\documentclass[10pt]{iopart}

\usepackage{graphicx}
\usepackage{color}
\usepackage{textgreek}

\newcommand{\dytio}{Dy$_2$Ti$_2$O$_7$}

\usepackage{iopams}

\begin{document}

\title{Low-temperature high-frequency dynamic magnetic susceptibility of classical spin-ice Dy$_2$Ti$_2$O$_7$}

\author{S.~Teknowijoyo$^{1,2}$, K.~Cho$^1$, E.~I.~Timmons$^{1,2}$, M.~A.~Tanatar$^{1,2}$, J.~W. Krizan$^3$, R.~J. Cava$^3$ and R.~Prozorov$^{1,2}$}

\address{$^1$ Ames Laboratory, Ames, IA 50011, USA\\
$^2$Department of Physics \& Astronomy, Iowa State University, Ames, IA 50011, USA \\
$^3$ Department of Chemistry, Princeton University, Princeton, New Jersey 08544, USA}

\ead{prozorov@ameslab.gov}
\vspace{10pt}


\begin{abstract}
Radio-frequency (14.6 MHz) AC magnetic susceptibility, $\chi^{\prime}_{AC}$, of \dytio\ was measured using self-oscillating tunnel-diode resonator. Measurements were made with the excitation AC field parallel to the superimposed DC magnetic field up 5 T in a wide temperature range from 50 mK to 100 K. At 14.6 MHz a known broad peak of $\chi^{\prime}_{AC}(T)$ from kHz - range audio-frequency measurements around 15~K for both [111] and [110] directions shifts to 45~K, continuing the Arrhenius activated behavior with the same activation energy barrier of $E_a \approx 230$~K. Magnetic field dependence of $\chi^{\prime}_{AC}$ along [111] reproduces previously reported low-temperature two-in-two-out to three-in-one-out spin configuration transition at about 1~T, and an intermediate phase between 1 and 1.5~T.   The boundaries of the intermediate phase show reasonable overlap with the literature data and connect at a critical endpoint of the first order transition line, suggesting that these low-temperature features are frequency independent. An unusual upturn of magnetic susceptibility at $T \to 0$ was observed in magnetic fields between 1.5~T and 2~T for both magnetic field directions,  before fully polarized configuration sets in above 2~T.
\end{abstract}

%
%
\submitto{\JPCM}
%
\maketitle
%
\ioptwocol

\section{Introduction}

Spin-ice compounds with a magnetic structure resembling the crystal structure of water ice have attracted notable interest since their identification in the 1990s \cite{Gaulin1992,Harris97,Ramirez99}. A rich variety of materials with the generic formula A$_2$B$_2$(O,F)$_7$ (here A is a trivalent cation and B is a tetravalent cation) with  pyrochlore lattice provides a fruitful playground for the search of exotic magnetic phases, including gapless spin liquid, quantum spin-ice, and magnetic monopoles, just to name a few \cite{Gingras2014}. In the classical spin-ice compounds, \dytio\ (DTO) and Ho$_2$Ti$_2$O$_7$ (HTO), the magnetic moments of the (Dy, Ho) rare earth elements are localized at the vertices of the corner-shared tetrahedra of the pyrochlore lattice \cite{Harris97,Ramirez99,Gingras2009}. Because of a strong single-ion anisotropy, the ground state of Dy$^{3+}$ is well expressed by an Ising doublet with local [111] quantization axes \cite{Rau2016}. An effective ferromagnetic nearest-neighbor interaction leads to a geometric frustration, which prevents the system from reaching a long-range ordered state \cite{Fukazawa02}. While the magnetic moments are macroscopically disordered, their local configurations obey the ``ice rule": two spins point outward and two spins point inward in each tetrahedron, which is analogous to the Bernal-Fowler rule for water ice \cite{BF}. As a result, the lowest energy state is degenerate with a residual entropy very close to that of water ice \cite{Ramirez99}. The resultant magnetically frustrated configuration is extremely sensitive to the competition between short-range and long-range magnetic interactions, leading to a complex magnetic behavior \cite{Gaulin1992,Harris97,Ramirez99,Fennell2005,Pomaranski13,Borzi2016}.

The extensive spin degeneracy is lifted when external magnetic field is applied. For a field along [100] direction, all four spins in an elementary tetrahedron align in a way such that they have a component of magnetic moment along the field direction, without breaking the ice rule \cite{SakakibaraPRL}. For a magnetic field along [111] direction, however, the response is more complex. For low fields, the moments first align following the ice rule so that one of the four spins on each tetrahedron is parallel to the field and forms a layer of triangular lattice which alternates with a layer of kagome spin lattice configuration \cite{SakakibaraPRL,Higashinaka2004}. When the applied magnetic field exceeds roughly 1~T, the phase with two-in-two-out (2-2) configuration of the spins undergoes a first order transition to a phase with three-in-one-out (3-1) configuration at low enough temperatures ($T\leq$ 0.36~K) \cite{SakakibaraPRL}.

Magnetization measurements with precisely aligned magnetic field direction found that a small deviation of a magnetic field from [111] direction towards [112] direction splits the 1~T transition into two \cite{Sato2007}. The resulting intermediate state was identified in specific heat \cite{Higashinaka2004} and susceptibility measurements \cite{Borzi2016,Sato2007,Grigera2015}. To explain this new state, a theoretical model taking into account the effects of dynamic lattice distortions has been proposed \cite{Borzi2016}.

In regards to the spin relaxation time, different spin configurations and transitions have very different characteristic time scale(s), from seemingly static to very slow activated behavior, to fast driven transitions. As a result, early studies of \dytio\ reported a variety of frequency dependencies of magnetic susceptibility, which showed at least two distinct peaks. The first peak at around 1~K is nearly frequency-independent at audio-frequency \cite{Matsuhira01} and its signature is also seen in zero-field heat capacity measurements \cite{Ramirez99,Pomaranski13}. At $T<1$~K, most of the relaxation channels ``freeze" out and the relaxation time scale grows exponentially on cooling. The possible dynamics of magnetic monopoles in this part of the phase diagram is an interesting suggestion, which has recently started to be explored experimentally \cite{Toews18,Stoter20}.
The second peak in AC susceptibility measurements exhibits strong temperature dependence, from  $\sim$0~K (DC) to 20~K (10~kHz AC) \cite{Matsuhira01,Snyder01,Fukazawa02,Snyder2004}. In the MHz range, a \textmu SR experiment found a gap in muon polarization between $40-60$~K \cite{Lago07} for which the authors attributed that it may or may not come from the same relaxation channel as found in the susceptibility studies. Measurements at even higher frequencies using nuclear forward scattering of synchrotron radiation \cite{Sutter} and nuclear magnetic resonance \cite{Kitagawa} also identify response even at higher temperatures.

Motivated by these studies and to bridge the gap between the audio frequency AC susceptibility and the higher frequency experiments, here we report radio-frequency (14.6 MHz) magnetic susceptibility study of the classic spin-ice compound Dy$_2$Ti$_2$O$_7$ using highly sensitive tunnel diode resonator (TDR) technique. Our zero field data show that the higher-temperature peak shifts to 45~K which is in a good agreement with the \textmu SR experiment \cite{Lago07}. Furthermore, we show that our peak temperature follows the same Arrhenius-type activated behavior as in low-frequency measurements. This observation supports the notion that the same energy barrier is responsible for this activated behavior across the wide frequency range. At low temperatures, we carefully explore the $H-T$ phase diagram down to $T_{\mathrm{base}}\approx 50$~mK and up to 2~T in both [110] and [111] directions. The frequency-insensitive 1~K peak grows stronger and we observe an intermediate phase near the spin flip transition for $H \parallel$[111]. We find that the magnetic action continues beyond the boundary of the intermediate phase before reaching fully polarized state above 2~T.

\section{Experimental}

Single crystals of \dytio\ were synthesized using optical floating zone method, as described in detail elsewhere \cite{Krizan2014}. The crystals were oriented using x-ray Laue diffractometry. The samples for the TDR measurements were cut and polished into the shape of cubes with sides of 0.7 mm long oriented along [110] or [111] directions. The accuracy of this cutting is estimated to be within 2 degrees from the determined direction \cite{Zeisner14}.

The tunnel diode resonator technique is widely used for precision radio frequency (rf) magnetic susceptibility measurements of magnetic and superconducting materials with $1:10^7$ frequency resolution measuring better than 0.1 Hz frequency shift at 10 MHz \cite{Prozorov00,Prozorov06,Vannette,VannetteJAP2008}. While in metallic samples the signal is a combination of skin effect and spin susceptibility, in insulating materials, such as the present case of spin-ice, it gives directly the real part of rf AC susceptibility up to a constant geometric factor. The TDR spectroscopy has been successfully used for studies of field-dependent phenomena and quantum magnetism, for example in magnetic molecules (e.g.Cr$_{12}$Cu$_2$) \cite{Engelhardt09,Martin09}, diluted magnetic systems \cite{KimRB66,Novikov2013} and FM/AFM materials \cite{Vannette,VannetteJAP2008,VannetteJMMM2008,Quirinale2017}. Please see reviews \cite{Prozorov00,Prozorov06,Prozorov11} for the detailed description of the principles of the TDR technique used in the present work, and Ref.~\cite{Kim2018RSI} for the technical discussion of the implementation of the TDR in a dilution refrigerator, which was used in our experiments in this work.

In brief, a tunnel-diode, biased to its regime of a negative differential resistance, drives an $LC-$ tank circuit consisting of a capacitor, $C$, and an inductor, $L$, with the fundamental frequency $2 \pi f_0 =(C L_0)^{-1/2}$ around 10-20 MHz. In our particular dilution refrigerator setup, $f_0 \approx $ 14.6 MHz. Essentially, the tunnel diode compensates for the losses and, when the impedances properly match, triggers spontaneous self-oscillations at $f_0$. It is possible to maintain this state with minimal external noise and excellent stability, provided that the circuit itself is well stabilized and shielded \cite{Kim2018RSI,VandeGrift75}. The outstanding sensitivity of the technique comes from the fact that it operates in the frequency domain (similar to microwave cavity perturbation technique and NMR), but without the need to scan the frequency (or field) in search for the resonance, which is an added benefit. A properly biased TDR system self-resonates and is always locked on its own resonance.
If a sample is inserted into the coil, the total (coil + sample) inductance changes from $L_0$ to $L=L_0+\Delta L$. In all practical setups, the inductance shift is small, $\Delta L \ll L_0$. The new resonant frequency is $2 \pi f = (C (L_0+\Delta L))^{-1/2} \approx f_0(1-\Delta L/2L_0+3/2(\Delta L/2L_0)^{2}+ \mathcal{O}(L/2L_0)^{3})$. Experimentally, the difference $\Delta f = f-f_0$ is in the order of 1~kHz. Since $f_0 \sim$ 10 MHz, $\Delta f/f_0 \approx 10^{-4}$ and therefore the quadratic and higher order corrections can be safely omitted.

The exact sign and value of $\Delta f$ are dependent on the susceptibility of the sample. A positive $\Delta f$ (i.e., negative $\Delta L$) implies a diamagnetic response and the opposite occurs for a paramagnetic one. As a function of temperature, the relation between the frequency change to the susceptibility can be written as $\Delta f(T)=-G4\pi\chi^{\prime}_{AC}(T)$. The constant of proportionality, $G=f_0V_s/2V_c(1-N)$, is determined by the demagnetization factor $N$, the sample volume $V_s$ and the coil volume $V_c$. This constant can be calibrated experimentally by measuring the absolute value of the DC susceptibility of the same sample in a Quantum Design Magnetic Properties Measurement System (MPMS). It can also be determined directly by pulling the sample out of the coil. The important requirement for the high resolution TDR measurements is the temperature stability of the coil and the circuit within a mK range. This is achieved by thermally separating the sample stage and the circuit parts as described in detail in \cite{Kim2018RSI}. The noise level of our TDR setup is $\sim$0.05 Hz out of 14.6~MHz, which means that the sensitivity of $\sim$3 ppb is achieved.

To cover an extended temperature range, we use two separate TDR setups in our study, one mounted in a $^3$He refrigerator fitted with a 9 T superconducting magnet (and equipped with a mechanical manipulator to perform sample pull-out) and another in a dilution refrigerator working with a 14 T magnet \cite{Kim2018RSI}. In both setups, the static field $H_{DC}$ is always parallel to the rf field $H_{AC}$ ($\sim$20 mOe). Measurements in the dilution refrigerator are performed at temperatures as low as 50 mK and up to 3~K. We therefore achieve a good temperature overlap with the $^3$He system used from 400~mK to 120~K. For \dytio, the low temperature region is of a particular interest because of the slowing down and freezing of the spin dynamics \cite{Gingras2009,Snyder01,Snyder2004} may allow the system to reach a magnetically ordered ground state in an unusual way.

DC magnetic susceptibility measurements were performed in {\it Quantum Design} MPMS SQUID magnetometer. These measurements were performed on the same samples as used in TDR study, and were used for calibration purposes for radio frequency measurements.

\section{Results}

\begin{figure}
	\includegraphics[width=8cm]{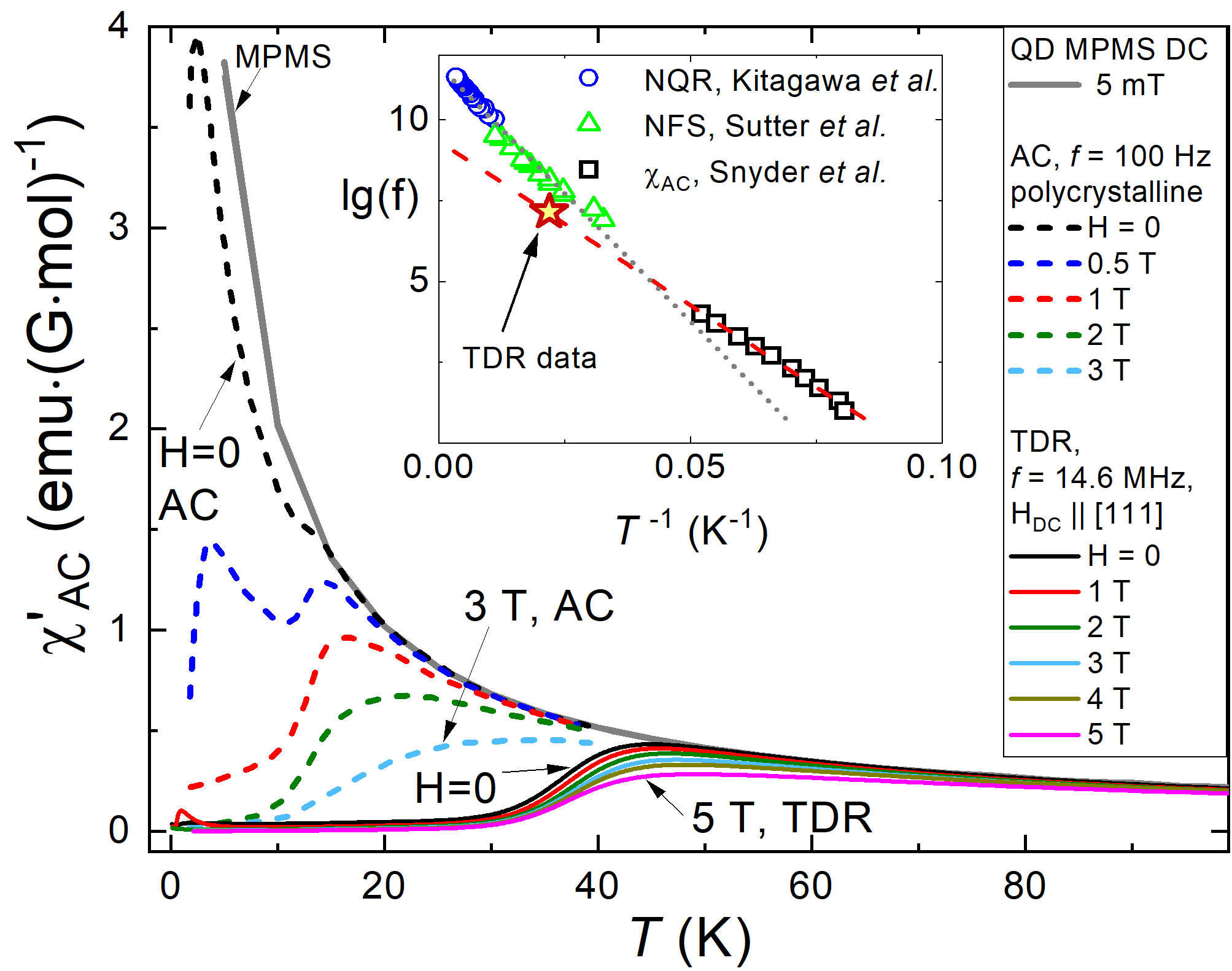}%
\centering
	\caption{(Color online) Temperature-dependent radio frequency (14.6 MHz) magnetic susceptibility $\chi^{\prime}_{AC}(T)$ of \dytio\ (solid lines) measured in magnetic fields $H\parallel$[111]. The gray solid line shows the $\chi^{\prime}_{DC}(T)$ of the same sample measured in a DC SQUID magnetometer for calibration. For reference, we show the 100~Hz $\chi^{\prime}_{AC}(T)$ data from Ref.~\cite{Snyder01} (dashed lines). The inset shows an Arrhenius plot of the position of the high temperature peak on the measurement frequency. Peak position in our RF measurements (red star with yellow interior) continues the trend determined in audio frequency measurements (open squares \cite{Snyder01}) with the same activation energy ($E_a$ $\approx$ 230~K, dashed line). The data from nuclear forward scattering (open magenta triangles) \cite{Sutter} and nuclear quadruple resonance (open blue circles) \cite{Kitagawa} experiments deviate slightly to a higher value of activation energy (dotted line).}
	\label{fig1}
\end{figure}

The solid lines in Figure~\ref{fig1} show the temperature-dependent magnetic susceptibility of \dytio\ measured at 14.6~MHz in magnetic fields up to 9~T applied along the [111] direction, using the $^3$He TDR setup. For reference, we show the $\chi^{\prime}_{AC}(T)$ data by Snyder \etal\ (dashed lines) measured at 100~Hz \cite{Snyder01}. The data for TDR spectroscopy are calibrated using the SQUID DC magnetic susceptibility measurements (gray solid line). The previously reported ``15~K feature" in low-frequency ($< 10$~kHz) susceptibility measurements \cite{Matsuhira01,Snyder01} shifts to $\sim 45$~K at 14.6~MHz and becomes less dependent on magnetic field. This temperature is also consistent with a previously reported $40-60$~K gap in the muon polarization in a similar frequency range \cite{Lago07}. As shown by the dashed line in the inset of Figure~\ref{fig1}, the position of the TDR peak (red star with yellow interior) extends the frequency dependence found in the low-frequency measurements \cite{Matsuhira01,Snyder01} almost in an exact linear fashion. The Arrhenius-type activation energy calculated using the formula $f = f_0$ exp$(E_a/k_BT)$ yields $E_a/k_B =$ (slope)/log$_{10}(e) \approx 230$~K and $f_0 \sim 2$~GHz. The value of the activation barrier is consistent with the theoretically calculated gap between the ground state and the first excited doublets in the spin-ice system \cite{Rosenkranz00} and therefore, the peak in $\chi^{\prime}_{AC}(T)$ has been attributed to the Orbach relaxation process between these states \cite{Lago07}. These mechanisms are also thought to be responsible for the results coming from  experiments at even higher frequencies such as nuclear forward scattering \cite{Sutter} and nuclear quadruple resonance \cite{Kitagawa}, which trace a  higher energy barrier with their combined average is shown as dotted line in the inset ($E_a/k_B\sim 365$~K). This difference could be related to the higher accessible excited crystal energy levels (due to different experimental parameters such as higher frequency and temperature), and the higher sensitivity to the local nuclear spin and quadruple interactions around the Dy$^{3+}$ ions of the of the NFS and NQR techniques.

\begin{figure}
	\includegraphics[width=8cm]{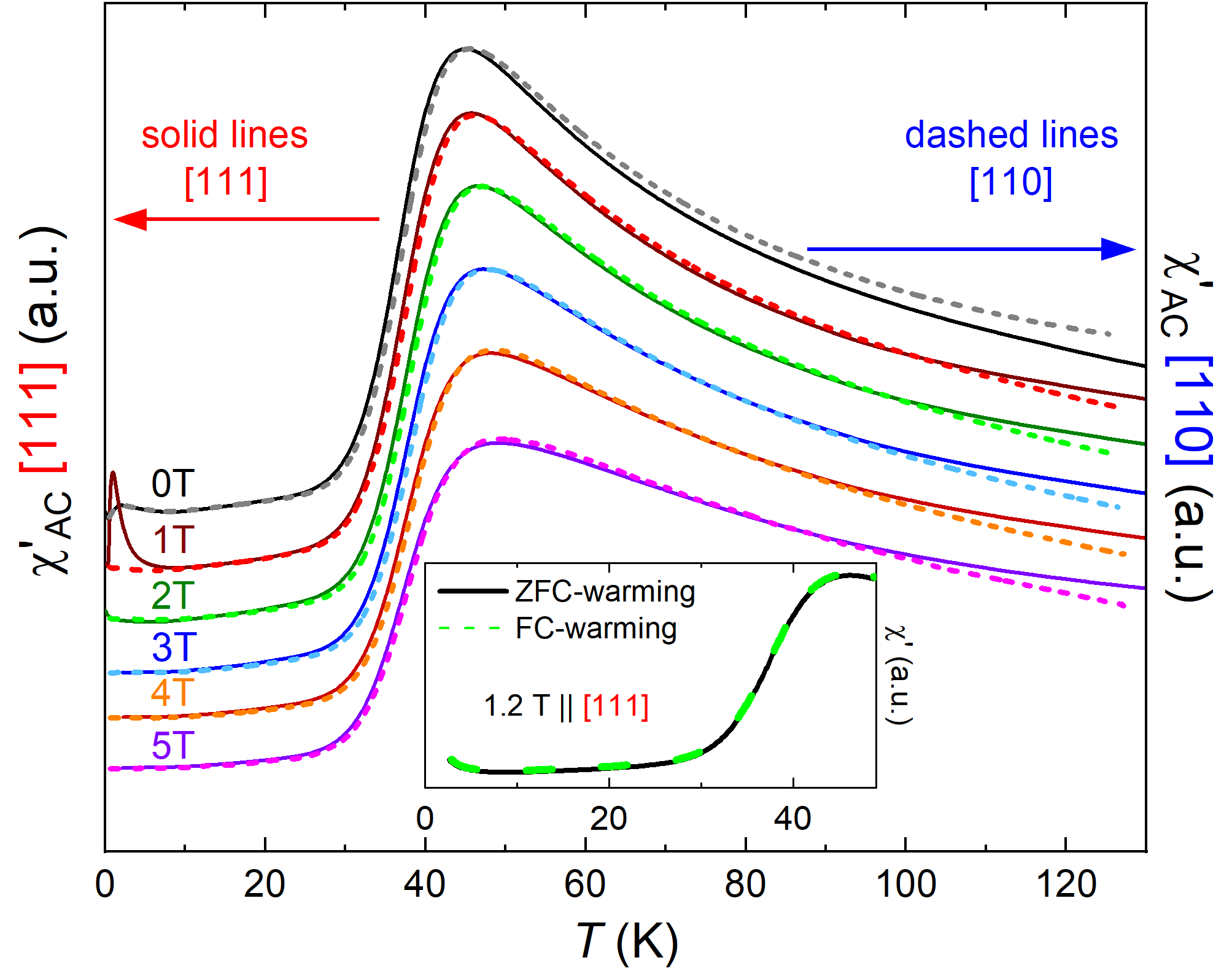}%
	\caption{(Color online) Temperature-dependent radio frequency (14.6 MHz) magnetic susceptibility $\chi^{\prime}_{AC}(T)$ of \dytio\ measured in magnetic fields $H\parallel$[111] (solid lines, left axis) and $H\parallel$[110] (dashed lines, right axis) directions. The curves for different applied magnetic fields are shifted vertically for clarity. For the sake of comparison of two independent measurements performed on crystals of different orientation, the right axis is adjusted so that $H \parallel$[110]=0 curve matches its [111] counterpart at the 45~K peak feature. The other dominant feature of $\chi^{\prime}_{AC}(T)$ is a sharp maximum observed at $\sim$1~K only in $H \parallel $[111] configuration in fields close to  1~T. The inset shows measurements on warming between 2.6~K to 50~K in $H \parallel$[111] = 1.2~T magnetic field for both ZFC (black) and FC (red) conditions. No obvious thermal hysteresis is found for both low-temperature and high temperature features. }
	\label{fig2}
\end{figure}

In Figure~\ref{fig2}, we compare TDR data for two experimental configurations, $H \parallel$[111] (solid lines, left axis)  and $H \parallel$[110] (dashes, right axis) over a broad temperature range $0.4-120$~K. The curves for different external DC magnetic fields are shifted vertically, from top to bottom, $|H|$ = 0, 1, 2, 3, 4 and 5 T.
The normalization factor between the two orientations is determined by matching the zero field curve for [110] to its [111] counterpart between 10 K and 50 K (around the Arrhenius-type spin relaxation peak, because its position and shape are nearly independent of field orientation and strength), and the same normalization factor is then applied to the rest of the [110] curves.
Above 50 K, the susceptibilities for the two orientations differ slightly presumably because of a small difference in the physical geometry of the two samples. A previous study on sample shape dependence of the magnetization measurements reported that spheres are highly desirable for the calculation of the exact (analytical) demagnetization correction \cite{Bovo2013}. For our cubic samples, the simple demagnetization correction $H_{i}\approx H-4\pi MN$ (where $H_{i}$ is the effective magnetic field sensed by the sample and $N$ is the geometric demagnetization factor) is no longer exact albeit still useful. This correction is applied for the region of our interest near the $\chi^{\prime}_{AC}(T)$ peak at $T\leq 1$~K in $H \parallel$[111] $\approx 1$~T (see Figure~\ref{fig6}(a) and the Discussion section below), which corresponds to a spin-flip transition from 2-in-2-out to 3-in-1-out configuration \cite{SakakibaraPRL,Aoki2004}.

Our data show no obvious thermal history dependence, as shown in the inset of Figure \ref{fig2}. Here we compare the data for measurements in $H \parallel$[111] $=1.2$~T taken for both zero-field cooled (the magnetic field is switched on after initial cool down at 2.6~K) and the consecutive in-field cooled run. The shown ZFC and FC curves are taken on warming with the same heating protocol in order to make their comparison straightforward.

\begin{figure}
	\includegraphics[width=8cm]{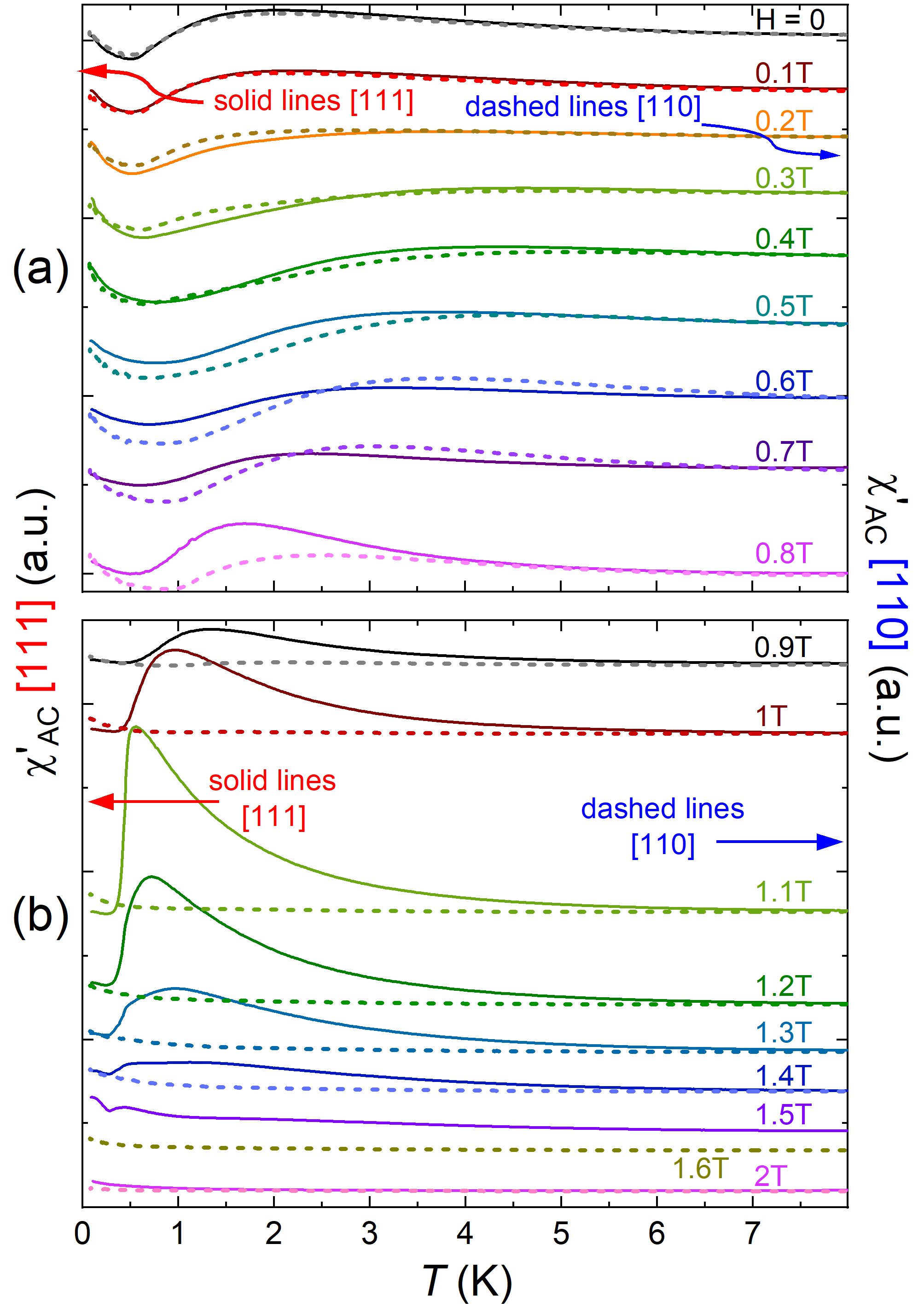}
	\caption{(Color online) Radio frequency magnetic susceptibility of \dytio\ measured in dilution refrigerator and $^3$He setups in 50 mK to 8~K temperature range. Measurements in configuration  $H \parallel$[111] are presented by solid lines with respect to the left axis, measurements in $H \parallel$[110] are shown with dashes with respect to the right axis. Panel (a) shows data for fields $H \leq $ 0.8~T and (b) for fields $H \geq $ 0.9~T in the vicinity of the spin-flip transition. Both left and right scales in (b) are $\sim$4 times larger to accommodate the height increase of the peaks. }
	\label{fig3}
\end{figure}

Figure \ref{fig3} shows in more detail the combined $\chi^{\prime}_{AC}(T)$ measurements from the dilution refrigerator setup (50~mK $-$ 3~K) and the $^3$He setup ($0.4-8$~K) in an external DC magnetic fields between $0- 2$ T. It has a similar construction to the Figure~\ref{fig2}, where the solid lines, left axis (dashed lines, right axis) correspond to the [111] ([110]) magnetic field direction and the curves are offset vertically for clarity.
The same normalization factor used in Figure~\ref{fig2} is applied again to the [110] curves for consistency in the comparison between [111] and [110] directions across different figures and panels.
Panel (a) shows data for fields below 0.8~T, where the difference between the two experimental configurations is not very pronounced. The zero-field curves for both configurations show a broad peak at around 2~K, followed by an increase of $\chi'$ response for $T \leq 0.5$~K. Both features are of a broad cross-over type, consistent with no observed long-range magnetic ordering down to 50~mK. Under increasing fields, the peak at 2~K broadens and both the peak and the onset of the upturn shift slightly to higher temperatures. The broad peak feature is similar to previous low-frequency susceptibility measurements \cite{Fukazawa02,Matsuhira01,Snyder01,Snyder2004} which is related to 2-2 to 3-1 spin flip transition. As $T\rightarrow 0$, the increase of susceptibility in almost all the curves in the figure bears similarity to the heat capacity data \cite{Pomaranski13} whose nature has been suggested to be related to the onset of phase transition to its (long-range ordered) ground state \cite{Pomaranski13}, Schottky-type peak due to multitude of local orders \cite{Borzi2016}, the dynamics of spin monopoles \cite{Toews18,Stoter20}, or the presence of other spin relaxation channels \cite{Shannon12,Melko2001}.

\begin{figure}
	\includegraphics[width=8cm]{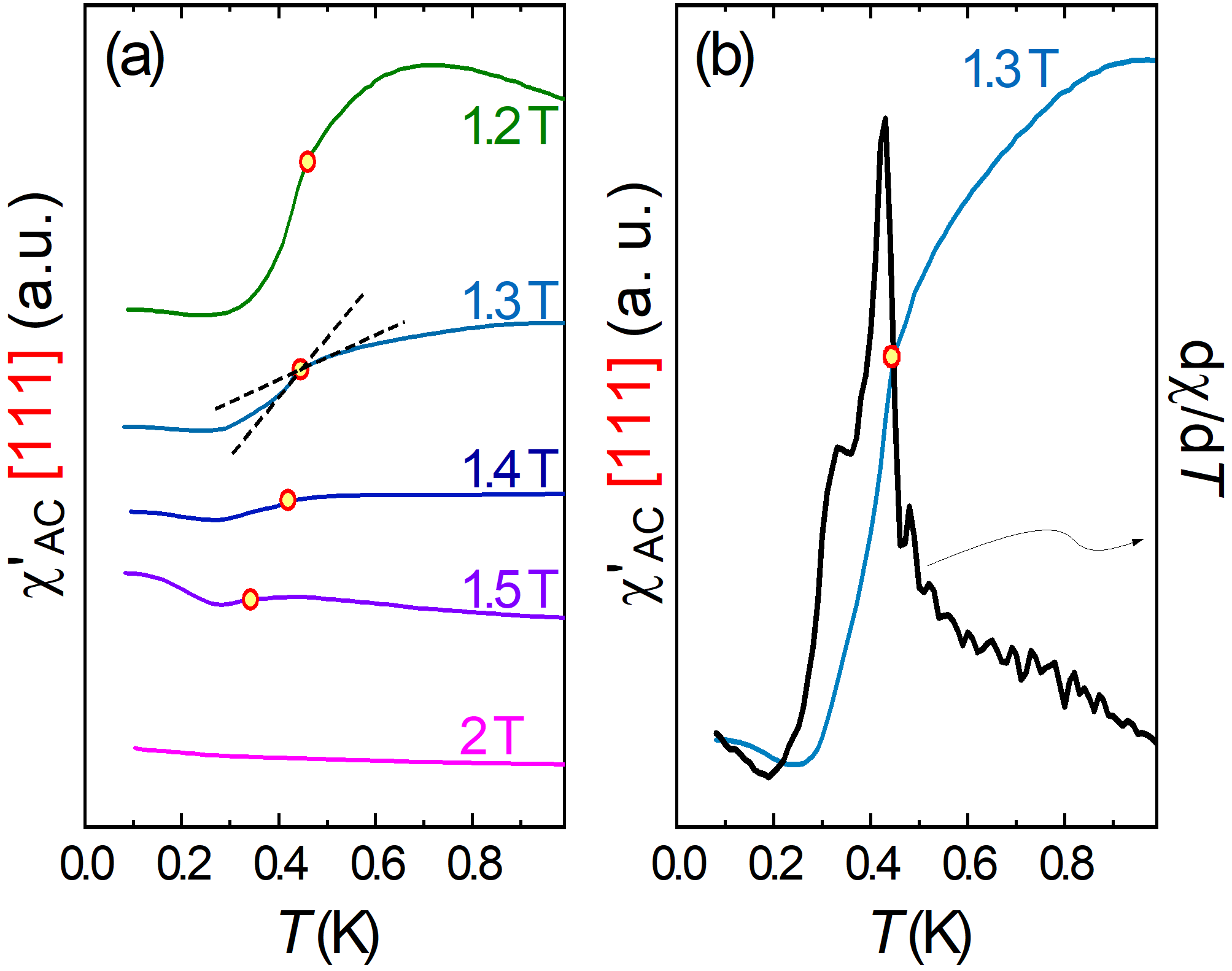}%
	\caption{(Color online) (a) Temperature-dependent real part of the radio frequency AC magnetic susceptibility of \dytio  $ $ taken in magnetic fields along the [111] direction in the $1.2 - 2$~T range. Non-monotonic features in the temperature dependencies are observed at $T \sim$ 0.4 K (circles), suggestive of an intermediate phase between $1.2 - 1.5$~T range which is absent in the 2 T curve. (b) An enlarged view of the susceptibility data in 1.3 T, emphasizing the ``kink" feature that is also clearly seen as a drop in the temperature-dependent susceptibility derivative (right axis).}
	\label{fig4}
\end{figure}

Figure~\ref{fig3}(b) shows measurements for $0.9~\mathrm{T}\leq |H| \leq 2$~T. In the vicinity of the spin-flip transition, $\chi^{\prime}_{AC}(T)$ in $H \parallel$[111] configuration shows a significant increase (note the curve at 1.1~T) and the build up of a strong peak at 0.7~K. The feature at 1.1~T is consistent with a step in the DC magnetization $M(H)$ measurements, as reported in the literature \cite{SakakibaraPRL,Aoki2004}. For fields of 1.2~T and above, the susceptibility gradually decreases, however, this decrease reveals some new sharp ``knee" features in the $\chi^{\prime}_{AC} (T)$ and its derivative (Figure~\ref{fig4}(a) and (b)), suggestive of a phase transition. None of these features are observed for the $H_{\mathrm{DC}} \parallel$[110]~
$\geq 1$~T configuration, for which the susceptibility shows a smooth evolution and remains small.
A specific heat Monte Carlo study has predicted a phase transition at low temperature when an external field is applied in the [110] direction \cite{Ruff2005} which is notably absent in our data and to the best of our knowledge, still eludes an experimental verification \cite{Grigera2015}.

\begin{figure}
	\includegraphics[width=8cm]{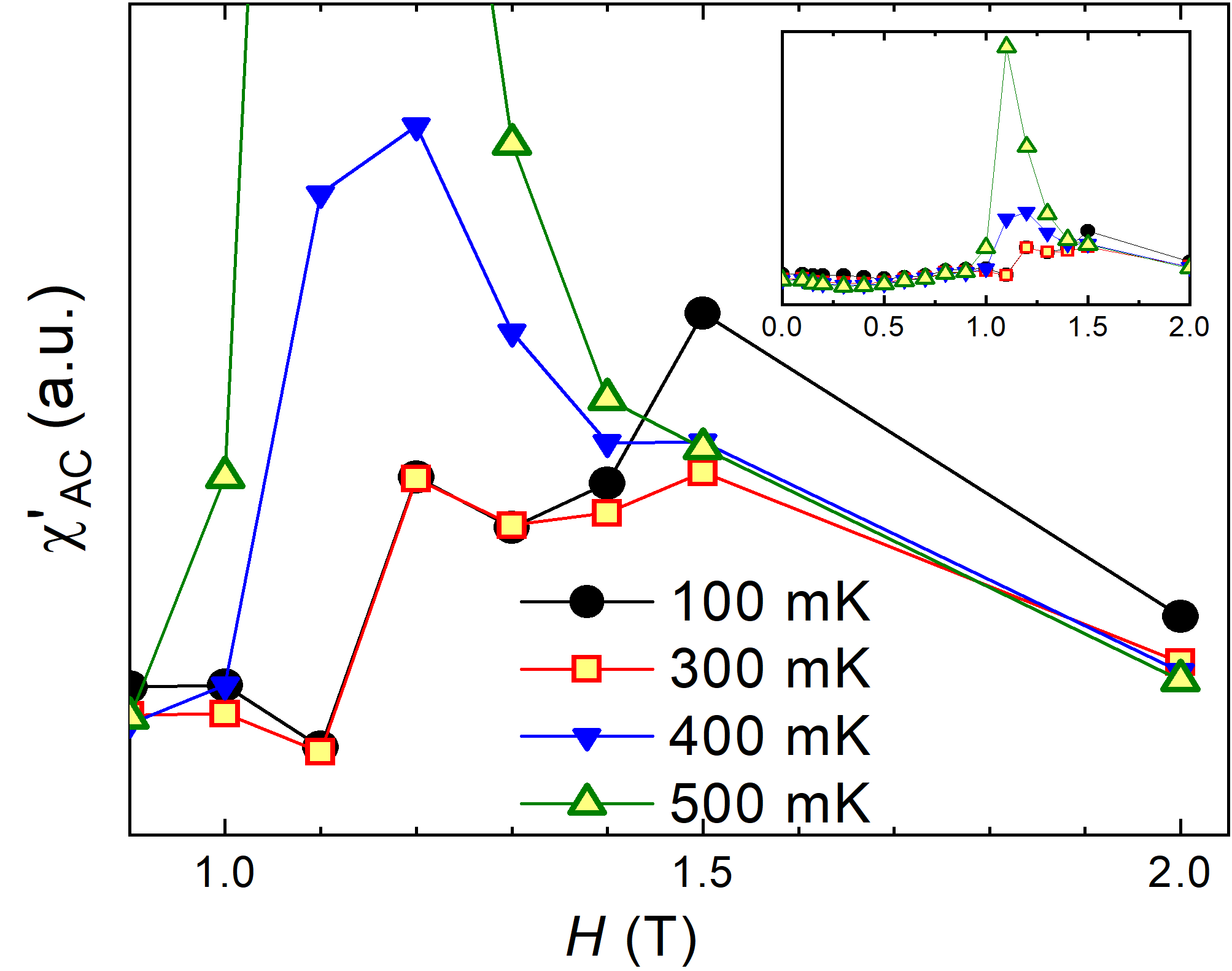}
	\caption{(Color online) Field-dependent radio frequency magnetic susceptibility as determined from $T$-sweep data of Figure~\ref{fig3} at characteristic temperatures of 100 mK (black solid circles), 300 mK (yellow-red squares), 400 mK (blue solid down-triangles) and 500 mK (green open up-triangles). The curve for 200 mK data is presented in Figure~\ref{fig6}(a). The locations of the peaks in susceptibility are marked as blue up-triangles in the phase diagram Figure~\ref{fig6}(b). At 500 mK the curve forms a single peak with stronger relative intensity as shown in the inset.}
	\label{figHsweep}
\end{figure}

To get an additional insight into the nature of the kink feature observed above 1.1~T, we plot in Figure~\ref{figHsweep} the temperature evolution of the field-dependent $\chi^{\prime}_{AC}(H)$ as extracted from $T$-sweep data of Figure~\ref{fig4}. The dependence shows a singular peak at 400 and 500~mK which loses intensity and splits into two peaks at 1.2~T and 1.5~T below 0.3~K. This evolution is similar to the two-peak features found in the low frequency AC susceptibility data \cite{Borzi2016,Grigera2015}, see below.

\section{Discussion}

Working with magnetic samples, even paramagnetic ones, there is always a
concern regarding the demagnetization correction. Most papers, however, do
not discuss this and it could be hard to follow their data to the actual magnetic
field sensed by the spins. Therefore any comparison between different sets
of data should be taken with this uncertainty in mind. In our case, we used
cube-shaped samples. Indeed, while true demagnetization correction is only valid
in ellipsoidal samples where magnetic induction is constant inside the sample, it can still be used in the effective, integral sense to relate a total magnetic moment of a given sample with the effective change of integral magnetic susceptibility, $dM/dH$, where $H$ is the applied magnetic field. The shape-dependent correction has been recently calculated for arbitrary-shaped samples, including most frequent in the research cuboidal shapes \cite{Prozorov2018}. Of course, the next best after sphere is a cube, which at least has isotropic corrections along three principal axes. This is the case in the present work.
The effective demagnetization factor for a cube along any of the primary
directions is $N=0.39$ \cite{Prozorov2018}. Note that the sum of the demagnetization factors, $\sum_i=0.39 \times 3=1.17 > 1$. This sum is equal 1 only for ellipsoids and this caused significant confusion in the literature trying to deal with non-ellipsoidal shapes. Using cgs notation, the
corrected magnetic field, ``seen" by the sample is $H_{i}=H-4\pi MN$, where $M=m/V$ is volume magnetization and $m$ is the measured magnetic moment in
emu=erg/G. In our case, the sample has dimensions of $0.74\times 0.70\times
0.70$ mm$^{3}$, so that its volume is $V=3.63\times 10^{-4}$ cm$^{3}$.
Taking DC magnetization measured in Quantum Design MPMS on this sample, the $%
M(H)$ curve saturates at $m_{\mathrm{sat}}=$ $0.22$~emu, roughly above $20$ kOe of the
applied field at 5~K. In the region of interest, at 12 kOe, it is already $m=
$ $0.188$~emu which is not far below $m_{\mathrm{sat}}$. This implies that we can use a simple approximation of $\mu H/k_{B}T \gg 1$ for higher fields and lower temperatures. Importantly, being
close to saturation means that the demagnetization correction amounts to a
simple practically temperature and field - independent constant shift and, therefore cannot contribute to any new features in our temperature - dependent measurements.
Using $m=0.188$~emu, the constant demagnetization correction is roughly, $4\pi m N/V\approx 2.5$ kG. For smaller magnetic fields and higher temperatures, this correction becomes smaller, because magnetization
decreases. Of course, for small fields, demagnetization correction will stay
temperature dependent down to the lowest temperatures, but it does not affect higher-field constant correction.

\begin{figure}
	\includegraphics[width=8cm]{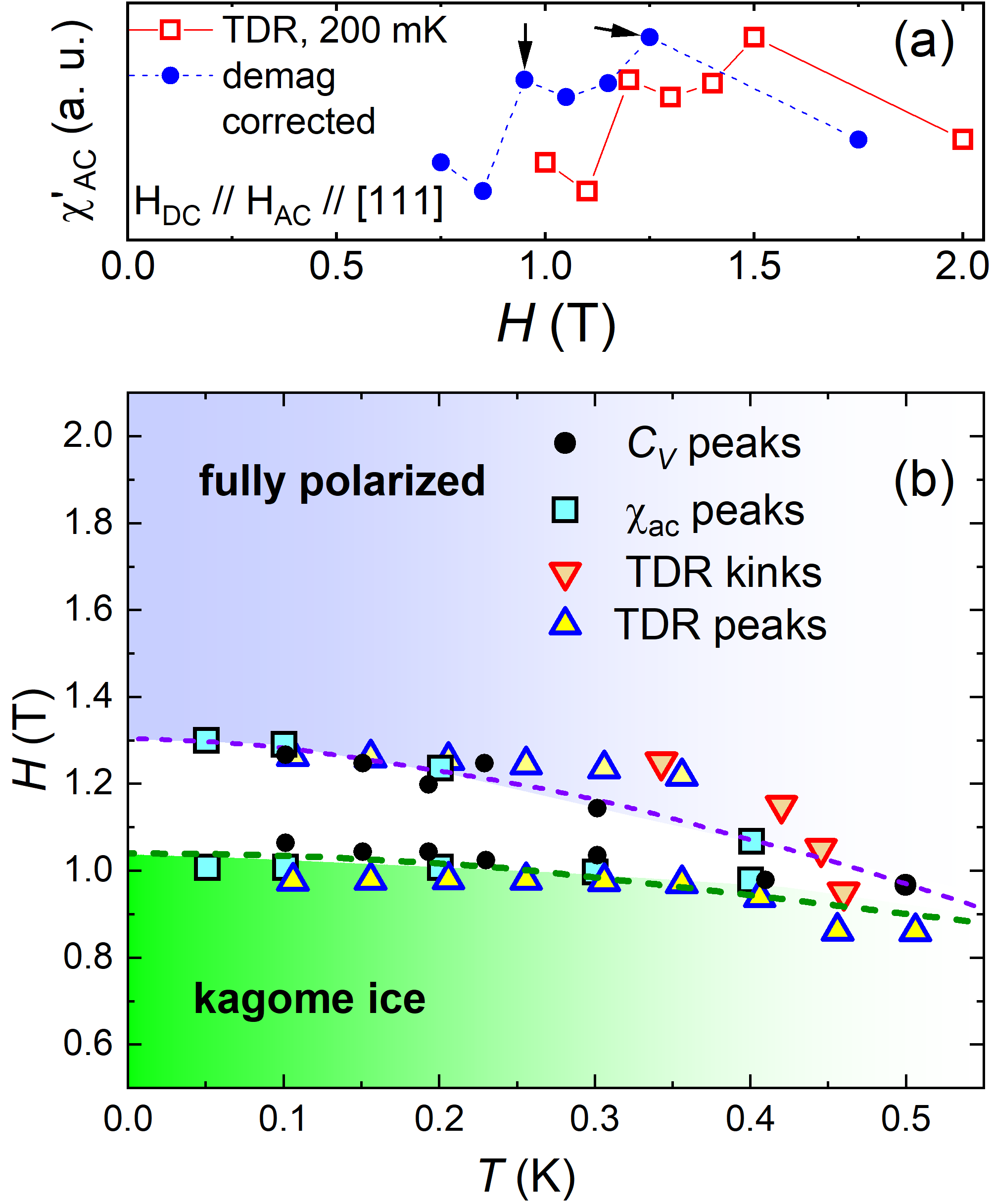}%
	\caption{(Color online) (a) Demagnetization correction of 0.25~T applied to the 200 mK data. The corrected peak locations are indicated by arrows. (b) Phase diagram of \dytio\ as determined from heat capacity (\cite{Higashinaka2004}, circles) and low-frequency AC magnetic susceptibility (\cite{Grigera2015}, squares) measurements in $H \parallel$ [111] configuration. These points form the spin-flip transition boundary (purple dashed line) and the kagom\'e-intermediate phase boundary (green dashed line). Red-yellow down-triangles denote the  kinks in the $T$-sweeps from Figure~\ref{fig4}a and and blue up-triangles denote the demagnetization corrected positions of the peaks of $H$-sweeps from Figure~\ref{figHsweep}.}
	\label{fig6}
\end{figure}

Figure~\ref{fig6} shows the demagnetization correction applied to the 200~mK data and the summary of the corrected special points in the $\chi^{\prime}_{AC} (T)$ and $\chi^{\prime}_{AC}(H)$ data. The red down-triangles represent the points of anomalies in the $T$-sweeps (Figure~\ref{fig4}), while the blue up-triangles are from the peaks in $H$-sweeps (Figure~\ref{figHsweep}). Note that the splitting of the critical point into two lines separated by approximately 0.25~T below 0.4~K is a characteristic feature of the intermediate phase as found in heat capacity (circles, from \cite{Higashinaka2004}) and low-frequency AC susceptibility data (squares, from \cite{Grigera2015}).
Our TDR data demonstrate reasonable overlap with these measurements which suggests that, unlike the 15~K feature, the peak splitting at low temperatures is largely frequency independent (up to the rf range).
This splitting feature is observed in magnetic fields slightly inclined from [111] toward [112] direction and the magnitude of the splitting depends on the angle of misalignment \cite{Borzi2016,Sato2007}.
The splitting in our data, $\sim$0.25~T, corresponds to inclinations between 5 and 7 angular degrees \cite{Sato2007} which is a reasonable estimate of the combined accuracy of our TDR coil-sample alignment and the sample crystallographic axes determination using Laue diffraction patterns.

Both our misalignment-induced splitting and the slight upturn at low temperatures are consistent with the distortion-based dipolar spin-ice model (d-DSM) \cite{Borzi2016}, which builds upon previous theoretical models \cite{Siddharthan1999,Hertog2000,Bramwell2001,Yavorskii2008}.
This slight increase of magnetic susceptibility on cooling is observed in magnetic fields of both [111] and [110] orientations, above 1.5~T boundary of the intermediate phase (Fig.~\ref{fig3}). It is generally believed that a spin polarized state is achieved above the intermediate phase, in which case rf susceptibility should be zero \cite{VannetteJAP2008}. Non-zero value suggest that the magnetic polarization is not complete, at least up to 2~T.



\section{Summary}

To summarize, we have extended the dynamic magnetic AC susceptibility measurements into the radio-frequency range by using the tunnel diode resonator technique to investigate a classical spin-ice compound Dy$_2$Ti$_2$O$_7$. We observe a peak in susceptibility at 45 K, adding a point that follows the linear behavior in the Arrhenius plot ($\log_{10} f$ vs. $1/T_{\mathrm{peak}}$) from the published low frequency data, thus indicating that a single activation barrier around 230~K for the Orbach processes related to the crystal field energy. This value is slightly lower than the barrier reported by yet higher frequency nuclear forward scattering \cite{Sutter} and nuclear quadruple resonance \cite{Kitagawa} experiments. The difference may be due to the difference in the experimental parameters, contributions of higher order terms and the local sensitivity of these probes. We also find that the AC susceptibility increases as $T$ goes below 0.5~K for all fields including above the spin-flip transition. This is inconsistent with the conventional spin-polarized state which could be an onset of a phase transition or unusual spin relaxation channels such as the proposed spin monopole dynamics.
``Kink-like" and peak-splitting features are visible in the susceptibility for a magnetic field along the [111] crystallographic direction in a 1.2 to 1.5 T range. After correcting for demagnetization effects, these points reveal a reasonable overlap with the boundaries found in previous specific heat and magnetic susceptibility studies on the $H-T$ phase diagram, suggesting that these features arise from frequency-insensitive mechanisms. Our results are consistent with the d-DSM model, which has been proposed as the underlying microscopical theory of the classical spin ice \dytio.

\section*{Acknowledgment}

We thank R. Flint and M.~J.~P.~Gingras for illuminating discussions. This research was supported by the U.S. Department of Energy, Office of Basic Energy Sciences, Materials Science and Engineering Division through the Ames Laboratory [S.T., K.C., E.I.T., M.A.T., R.P.]. The Ames Laboratory is operated for the U.S. Department of Energy by Iowa State University under Contract No. DE-AC02-07CH11358. Work at Princeton University [J.W.K., R.J.C.] was supported by DOE BES grant number DE-SC0019331.

\section*{References}

\bibliographystyle{iopart-num}

\providecommand{\newblock}{}

\end{document}